# Explicit relationship between electrical and topological degradation of polymer-supported metal films subjected to mechanical loading

O. Glushko[1]*, P. Kraker[1], M.J. Cordill[1,2]

[1]Erich Schmid Institute of Materials Science, Austrian Academy of Sciences, Jahnstrasse 12, A-8700 Leoben, Austria

[2]Department of Material Physics, Montanuniversität Leoben, Jahnstrasse 12, A-8700 Leoben, Austria

*corresponding author's e-mail: oleksandr.glushko@oeaw.ac.at

**Abstract**

For a comprehensive characterization of mechanical reliability of metallization layers on polymer substrates both electrical and mechanical degradation should be taken into account. Although it is evident that cracking of a conductive film should lead to electrical degradation, the quantitative relationship between the growth of electric resistance and parameters of the induced crack pattern has remained thus far unexplored. With the help of finite element modelling we were able to find an explicit and concise expression which shows that electrical resistance grows with the fourth order of the crack length and second order of the areal crack density. The discovered relationship was verified by comparison with the experimental results of tensile testing of polymer-supported thin metal films. Presented model is independent of the length scale and can be applied to films with different thicknesses as long as Ohm's law is valid. It is demonstrated that linear crack density is an ambiguous parameter which does not properly capture the development of a crack pattern. For the unambiguous characterization of the intensity of a crack pattern a universal dimensionless factor is proposed. Presented results show that there is a wide range of possible crack patterns which do not lead to electrical failure of a conductive film that can be used for failure-free design of flexible electronic devices.

Recently a great deal of attention has been focused on the investigations of mechanical reliability of polymer-supported thin conductive films. The main driving force for this research area is rapid development of flexible electronics concepts and technologies [1,2]. The most straightforward way to characterize the evolution of mechanical damage is to use optical or scanning electron microscopy to track the structural changes at different stages of mechanical loading [3-8]. Additionally, one can record the electrical resistance of a film during mechanical test and correlate the changes in resistance signal to the structural changes within the film under test [3-17]. Significant structural degradation, such as cracking of a brittle film under monotonic loading, is indicated by a rapid growth of electric resistance [3-5]. In the case of ductile films, such as copper, silver or gold, the resistance increases slowly with applied strain and a reference curve which represents the constant volume approximation [7-11] is usually used to estimate the induced mechanical damage. Under cyclic mechanical loading the growth of resistance is attributed to continuous crack propagation with increasing cycle number [12-17]. Although the electro-mechanical testing approach was extensively used during last decade, the exact relationship between the growth of electric resistance and parameters of the crack pattern is still unknown. For brittle indium tin oxide (ITO) films the model assuming the existence of conductive bridges within the cracks was developed [3,5]. However, it is difficult to estimate the range of applicability of this model since the nature of the bridges was not explained.

In this paper, finite element analysis is used to simulate the flow of electric current in a conductive material sheet with different crack patterns. The results of simulations gave rise to an explicit relationship connecting the growth of electric resistance with the parameters of the crack pattern. The proposed model is verified by direct comparison with experimental data obtained from monotonic tensile testing of thin polymer-supported films with in-situ resistance measurements.

Commercial finite element analysis software LISA 8.0.0 [18] was used to simulate the flow of electric current in a thin film containing cracks. The quasi-3D sheet of conductive material is described by the length ($L$), width ($W$), thickness, and conductivity. The crack pattern is defined by the length of a single crack, $l_0$, and total number of cracks, $N$. All cracks are oriented perpendicular to the direction of electric current and have the same length. Areal crack density, $C_a$, is defined as a number of cracks per area and can be expressed as

$$C_a = \frac{N}{LW}. \tag{1}$$

Linear crack density, $C_l$, is defined as a number of cracks per length in the current direction and is given by

$$C_l = \frac{Nl_0}{LW} = C_a l_0 . \tag{2}$$

The following simulation procedure was utilized. First, the global parameters of the material sheet, such as *L*, *W*, thickness, and conductivity are defined. The solution of the model without cracks gives the drop of electric potential along the material sheet which is used to calculate the initial resistance by means of Ohm's law. Then, the values of the length of a single crack, $l_0$, and the total number of cracks, *N*, are chosen. The cracks are introduced in the material sheet by selecting finite elements in the model and turning their conductivity to zero. The positions of the cracks are randomly generated using an external script. Five random crack configurations are generated for each pair of {*N*, $l_0$} for statistics. In this way, the average resistance of the conductor sheet with a random crack pattern characterized by particular values {*N*, $l_0$} is calculated. An example of simulated material sheet with random crack configuration is given in Supplementary Material.

Since the considered model is not restricted to any particular length scale the arbitrary length units "l.u." will be used in order to keep the generality of the solution. The dependence of the relative resistance on the linear crack density for different values of $l_0$ calculated for *L*=200 l.u. and *W*=50 l.u. is shown in Fig. 1 by the solid square symbols connected by solid lines. The same calculations were repeated for different values of *L* and *W*. Under the requirement that the length of a single crack is sufficiently smaller than the sample width, no statistically significant effect of model dimensions on the results was found.

It follows from Fig. 1 that the growth of resistance due to cracks exhibit a strong dependence on both the linear crack density, $C_l$, and length of a single crack, $l_0$. By fitting the curves obtained from simulations we were able to find a simple and explicit formula for the relative resistance:

$$\frac{R}{R_0} = 1 + \frac{1}{\sqrt{2}} C_l l_0 + \frac{1}{2} C_l^2 l_0^2 . \tag{3}$$

The dashed curves in Fig. 1 show Eq. (3) calculated for the same parameters as the corresponding simulation results. One should notice the excellent agreement between Eq. (3) and the numerical results. The deviations become significant only at high densities ($C_l > 0.4$ l.u.$^{-1}$) of long cracks ($l_0$=10 l.u.). This is explained by the increasing probability of random patterns where several cracks are merged together.

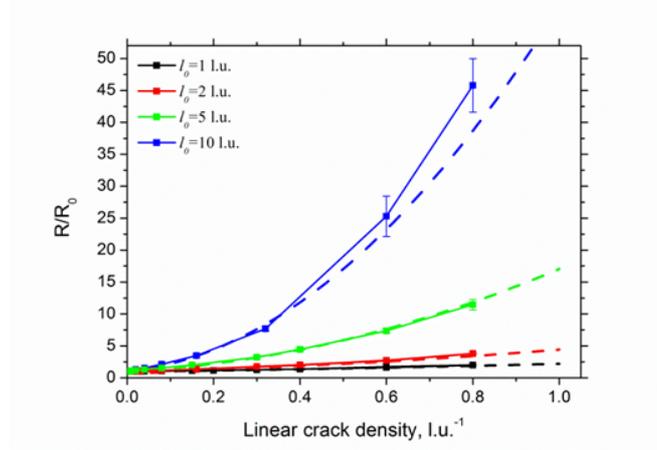

Fig. 1. The dependence of relative resistance on the linear crack density for different crack lengths, $l_0$. Solid squares connected by solid lines show the results of the numerical simulations with standard deviations over five random configurations. Corresponding dashed lines are obtained from Eq. (3). Arbitrary length units (l.u.) are used to keep the generality of the solution.

Equation (3) has a symmetric form with respect to two variables ($C_l$, $l_0$) and is considered to be the most appropriate for the characterization of real samples since the linear crack density is the typical experimental parameter used for the description of cracking in polymer-supported films [3-7]. However, since $C_l$ depends on $l_0$, a more rigorous expression of resistance as a function of two independent variables would be

$$\frac{R}{R_0} = 1 + \frac{1}{\sqrt{2}} C_a l_0^2 + \frac{1}{2} C_a^2 l_0^4. \qquad (4)$$

Equation (4) shows that the electrical resistance grows with the fourth order of the crack length and second order of the areal crack density. It is necessary to note that Eqs. (3) and (4) do not contain any sample dimensions and are only valid if the crack lengths are much smaller than the sample width.

For experimental verification of Eq. (3), three different thin film systems were chosen and strained to 20% to induce cracks. The resistance was recorded in-situ during loading, unloading as well as for a further 2 hours after unloading in order to account for viscoelastic relaxation of the substrate [8]. The average crack length and linear crack density were measured for each sample type using scanning electron micrographs. The chosen thin film systems exhibit distinctly different crack patterns, as shown in Fig. 2. Inkjet printed silver films with the thickness of 800 nm on polyethylene naphthalate (PEN) which are referred below as T1 (Fig. 2a) exhibit long cracks with the lengths of 9.1±3.1 µm and a low crack density of 0.17±0.02 µm$^{-1}$. Sample type T2 (Fig. 2b) corresponds to inkjet printed 400 nm thick printed silver films on polyethylene terephthalate (PET) having short cracks with the lengths of 2.1±0.9 µm and a high crack density of 0.51±0.05 µm$^{-1}$. Finally, thermally evaporated 100 nm thick silver films on PEN substrate which are defined as T3 (Fig. 2c) have short cracks with the lengths of 1.85±0.81 µm and a relatively low crack density of 0.23±0.04 µm$^{-1}$

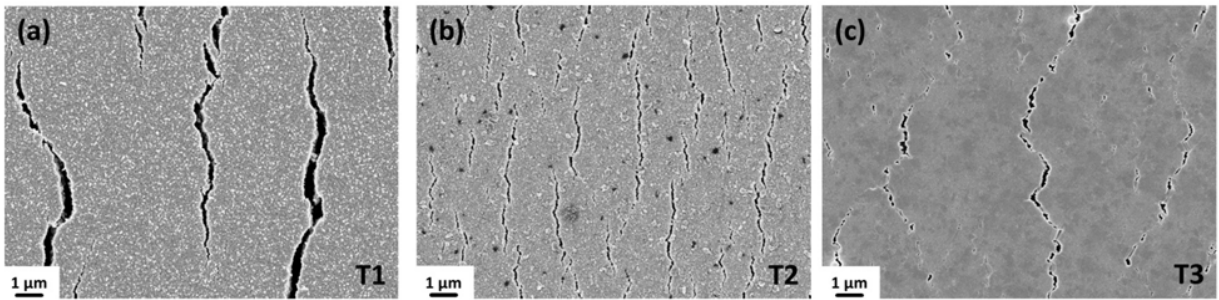

Fig. 2. Scanning electron micrographs of the three types of thin film systems after straining to 20%. (a) 800 nm thick printed silver on PEN, (b) 400 nm thick printed silver on PET, (c) 100 nm thick evaporated Ag on PEN.

The comparison of the experimental and simulated resistance growth is visualized in Fig. 3. The dashed lines depict Eq. (3) calculated for different values of $l_0$ which are specified on the plot. The unit length is now set to 1 µm for the sake of consistency with the experimentally observed crack patterns. For illustrative purposes the hatched areas depict approximate deviations of experimentally measured crack lengths and crack densities for the three considered sample types (T1, T2, and T3). The green circles show the values of the resistance calculated using Eq. (3) with the experimentally measured mean values of $C_l$ and $l_0$. The red square symbols depict the corresponding values of final relative resistance measured experimentally. The deviation between the calculated and experimental values of resistance growth is very low for T3 being higher for T2 and T1. However, keeping in mind the relatively high standard deviations of the experimentally measured crack lengths, the correspondence between the theoretical and experimental values is considered to be good.

A very important conclusion which can be drawn from Fig. 3 is that even high crack densities can result in only a moderate resistance increase if the lengths of the cracks can be kept small. It is also clear that neither crack density nor crack length alone could provide an unambiguous description of the damage development. Thus, the widely used practice of using only linear crack density to correlate induced mechanical damage with the resistance growth [3-7] should be corrected.

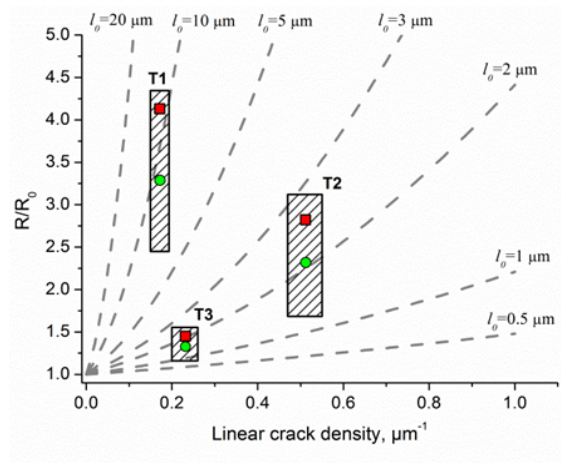

Fig. 3. Comparison of the experimental and theoretical data. The dashed lines are obtained from Eq. 3 for different values of $l_0$ which are specified on the plot. The hatched areas represent the approximate spread of experimental linear crack densities, $C_l$, and crack lengths, $l_0$, for three samples types which are shown in Fig. 2. The green solid circles are the mean values of electric resistance calculated from Eq. 3. The red solid squares depict the corresponding experimentally measured resistance values.

The crack length and crack density can be combined in a single variable describing mechanical degradation of a thin film. This new variable which we call the cracking factor, $C_F$, is defined simply as a product of the crack length and linear crack density: $C_F = C_l l_0$. Keeping in mind that linear crack density can be considered as a reciprocal of the average crack spacing, $\lambda$, the cracking factor can be also represented as $C_F = l_0/\lambda$. To support the implementation of the cracking factor consider two idealized crack patterns shown in Fig. 4. In Fig. 4a there are four cracks within the square area $D \times D$ ($D$ is much smaller than the width of the sample) and each crack has a length of $l_0 = D/2$. According to Eqs. (1) and (2), the areal crack density, linear crack density, and cracking factor are $4/D^2$, $2/D$, and 1, respectively. In Fig. 4b there are two cracks with the length $l_0 = D$ each within the same area. The areal crack density, linear crack density, and cracking factor are $2/D^2$, $2/D$, and 2, respectively. The situation shown in Fig. 4b can be considered as the result of coalescence of the cracks shown in Fig. 4a. One can see that despite the more intensive crack pattern the areal crack density becomes even lower in comparison to the case shown in Fig. 4a. The linear crack density stays the same and thus does not reflect the fact that the crack length increased. The cracking factor in Fig. 4b is two times higher than in Fig. 4a and is considered to be more appropriate measure of the cracking intensity. The fact that the cracking factor is dimensionless allows it to be applied at different length scales as a universal measure of cracking intensity.

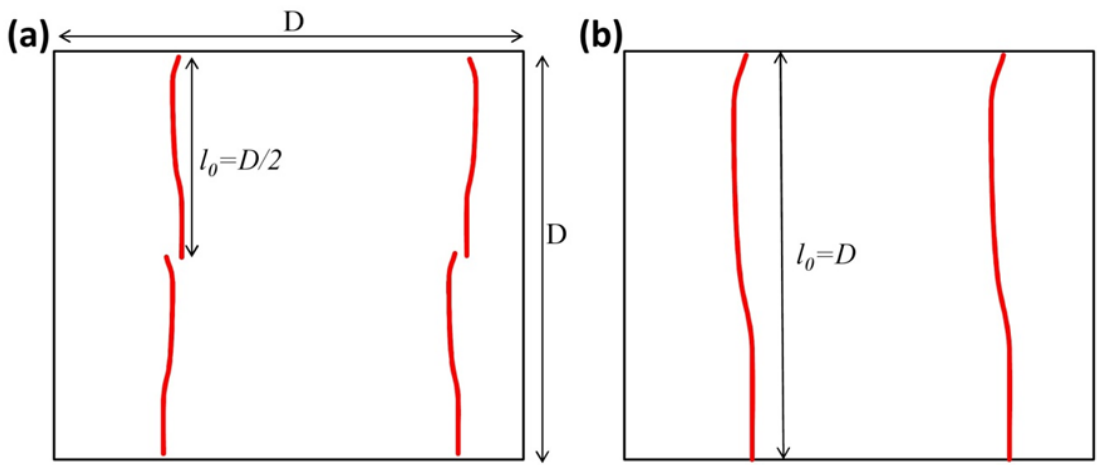

Fig. 4. An example of two idealized crack patterns with different intensities. (a) Four cracks within the area $D \times D$, each crack has a length of $l_0 = D/2$. (b) Two cracks with the length $l_0 = D$ within the same area.

Using the cracking factor, one can now provide a two-variable equation connecting the electrical and topological degradation of conductive films:

$$1 - \frac{R}{R_0} + \frac{1}{\sqrt{2}} C_F + \frac{1}{2} C_F^2 = 0 . \qquad (5)$$

With Eq. (5) the intensity of cracking can be determined if the growth of resistance is known or the electrical degradation of the film can be calculated if the cracking factor is known. Thus, the electrical degradation can be estimated without any electrical measurements. For instance, the crack patterns shown in Fig. 4a and Fig. 4b should result in resistance growth of $R/R_0 \approx 2.2$ and $R/R_0 \approx 6.4$, respectively. On the other hand, if there is a defined tolerance value for electric degradation for a certain thin film system, then a corresponding tolerance cracking factor can be calculated from Eq. (5). For example, to keep the growth of resistance below 20% the cracking factor should stay below 0.242.

The presented model, as any mathematical model, has its applicability range and limitations. The main model assumptions which should be kept in mind when applying it are listed below.

1. All cracks are assumed to be through-thickness cracks. This assumption seems to be legitimate since only a though-thickness crack provides a clear barrier for electric current. An in-situ AFM study of thin copper films under monotonic loading showed that the resistance starts to grow first when through-thickness cracks develop [20]. It was also experimentally shown that local thinning and strong surface roughening of a thin gold film under fatigue loading conditions does not cause a significant resistance increase if through-thickness cracks are not observed [15].

2. The presented results are applicable for one-dimensional crack patterns where all cracks are running virtually parallel to each other and perpendicular to the direction of electric current. Such crack configuration is a typical response to uniaxial tensile strain. The two-dimensional interconnected crack patterns which are usually observed in the case of biaxial strains are out of scope of this model.

3. All cracks are assumed to be totally insulating with no current flowing across the crack. In reality the cracks formed at the maximum applied strain can partially restore their conductivity due to re-connection during unloading when a visco-elastic substrate imposes compressive strains within the film. The effect of this resistance recovery was systematically investigated previously [8]. In the case of crack re-connection the model can still be used if every open portion of a larger crack is considered as a separate crack.

4. The cracks lengths should be significantly smaller than the sample width. It is a clear limitation since the sample width is not included in the Eqs. (3-5). However, the finite element model does not have this limitation and can also give an open circuit solution (infinite resistance) if a single crack spans across the sample width.

In summary, a finite element model for the resistance growth in a quasi 3D material sheet with random one dimensional crack patterns was developed and solved for different crack configurations. By fitting the simulation results an explicit expression for the resistance growth was found and verified by comparison with experimentally measured data. A dimensionless factor, the cracking factor, is introduced and shown to be a universal parameter which appropriately describes the intensity of cracking of a thin film on a polymer substrate. The explicit relationship provides a unique possibility to estimate mechanical damage knowing only electrical resistance, or alternatively the electrical degradation can be estimated on the basis of only microscopic characterization.

**Supplementary Material**

See supplementary material for the example of a finite element model of a conductive material sheet with random crack pattern.

**Acknowledgements**

This work was supported by the Austrian Science Fund (FWF) through project P27432-N20. The authors would like to thank C. Kirchlechner for valuable discussions.